\documentclass[10pt,conference]{IEEEtran}
\IEEEoverridecommandlockouts

\usepackage{cite}
\usepackage{graphicx}
\usepackage{amsmath,amssymb,amsfonts}
\usepackage{textcomp}
\def\BibTeX{{\rm B\kern-.05em{\sc i\kern-.025em b}\kern-.08em
    T\kern-.1667em\lower.7ex\hbox{E}\kern-.125emX}}
\usepackage{pifont} 
\usepackage{xcolor} 
\usepackage{bm}
\usepackage{array}
\usepackage{balance}

\usepackage{colortbl}
\usepackage{xspace}
\usepackage[framemethod=TikZ]{mdframed}
\usepackage{hyperref}

\usepackage{textcomp}
\usepackage{stfloats}
\usepackage{multirow}
\usepackage{url}
\usepackage{verbatim}
\usepackage[most]{tcolorbox}
\tcbuselibrary{breakable}
\usepackage{caption}
\usepackage{multirow}
\usepackage{booktabs}
\usepackage{calc}
\usepackage{bbding}
\usepackage{threeparttable}
\usepackage{algorithm}
\usepackage{algpseudocode}
\usepackage{subfigure}
\usepackage{float}
\usepackage{listings}
\usepackage{circledsteps}

\DeclareRobustCommand*{\IEEEauthorrefmark}[1]{%
  \raisebox{0pt}[0pt][0pt]{\textsuperscript{\footnotesize\ensuremath{#1}}}}

\newcommand{\ourtool}{\textsc{SpecFirst}\xspace}
\newcommand{\baselineDirect}{Direct-Synthesis\xspace}

\newcommand*\circled[1]{\tikz[baseline=(char.base)]{
            \node[shape=circle,fill,inner sep=1pt,font=\footnotesize] (char) {\textcolor{white}{#1}};}}

\begin{document}

\title{SpecFirst: Behavioral Specification
Elicitation as a First-Class Step in Agent-Based Program Synthesis from Scratch}

\author{
\IEEEauthorblockN{
Yihao Chen\IEEEauthorrefmark{2},
Shi Chang\IEEEauthorrefmark{1},
Khaled Chawa\IEEEauthorrefmark{4},
Feng Lin\IEEEauthorrefmark{1},
Boyuan Chen\IEEEauthorrefmark{1},
Shaowei Wang\IEEEauthorrefmark{3},
Ahmed E. Hassan\IEEEauthorrefmark{2}
}

\IEEEauthorblockA{
\IEEEauthorrefmark{1}Huawei Canada\\
\{shi.chang, feng.lin, boyuan.chen\}@huawei.com
}

\IEEEauthorblockA{
\IEEEauthorrefmark{2}Queen's University\\
yihao.chen@queensu.ca,
ahmed@cs.queensu.ca
}

\IEEEauthorblockA{
\IEEEauthorrefmark{3}University of Manitoba\\
shaowei.wang@umanitoba.ca
}

\IEEEauthorblockA{
\IEEEauthorrefmark{4}Concordia University\\
khaled.chawa@mail.concordia.ca
}
}

\maketitle

\begin{abstract}
LLM-based agents excel at software engineering tasks where an existing codebase provides context, but constructing a program from scratch remains fundamentally harder. Recent benchmarks such as ProgramBench quantify this gap: given only natural-language documentation and an execute-only binary as a behavioral oracle, even frontier models solve fewer than 1\% of instances. Existing frameworks conflate documentation reading, behavioral exploration, and code synthesis into a single pass, causing agents to probe insufficiently, lose behavioral intent as context drifts, and propagate early misinterpretations into the final implementation. Inspired by classical requirements engineering, we argue that behavioral specification elicitation should be a first-class phase that precedes implementation. We present \ourtool, a two-stage framework that forces the specification elicitation before code synthesis. A dedicated spec agent first probes the binary and combines observations with documentation into a structured specification. Next, a code synthesis agent then uses this specification to drive implementation. This decomposition resolves documentation ambiguities before coding begins and provides a stable behavioral reference throughout synthesis. We evaluate \ourtool on all 200 ProgramBench instances across four models spanning two families and an order of magnitude of capability. \ourtool consistently outperforms the single-loop baseline, improving test pass rates by 6.9\%--21.3\% and binary exploration coverage by 9.4\%--18.5\%, all statistically significant. Behavioral analysis on code synthesis further shows that a prior specification enables earlier and more sustained code construction. Our results demonstrate that an explicit requirements-engineering phase is an effective paradigm for from-scratch program construction.
\end{abstract}


\section{Introduction}
\label{sec:intro}

LLM-based agents have demonstrated strong performance on software engineering tasks such as bug fixing~\cite{jimenez2024swebench}, feature implementation~\cite{yang2024swe}, and code completion~\cite{chen2021codex,github_copilot}. Those tasks share a common structure: an existing codebase anchors the agent's work. Building a program entirely from scratch is fundamentally harder: the agent must decide everything (e.g., architecture and programming language) and reconstruct the full behavioral envelope of a target program from limited information alone. \textsc{ProgramBench}~\cite{yang2026programbench} makes this concrete: given only natural-language documentation and an execute-only binary as a behavioral oracle, an agent must produce a faithful re-implementation from scratch. Even the strongest frontier models (e.g., GPT-5.5) can only fully resolve fewer than 1\% of instances, confirming that from-scratch program construction remains an open challenge.

Existing agent frameworks such as SWE-agent~\cite{yang2024swe} and OpenHands~\cite{wang2025openhands} approach this task by providing the agent with natural-language documentation as the primary specification and an execute-only binary as a behavioral oracle~\cite{yang2026programbench}. While this setup permits behavioral discovery through probing, these frameworks \emph{mix} specification elicitation and code synthesis into a single undifferentiated loop---the agent switches freely between planning, reading documentation, probing the binary, and writing code within the same turn budget, with no mechanism forcing comprehensive elicitation before implementation begins. This entanglement leads to three systematic limitations.

\noindent\textbf{\circled{1} Ineffective exploratory probing.} Without a dedicated phase for behavioral discovery, the agent rarely probes the binary thoroughly enough to construct a complete behavioral model before committing to implementation. Edge cases, error paths, and flag interactions that are absent from the documentation but present in the binary are frequently missed, because the agent shifts to coding before its exploration is complete.

\noindent\textbf{\circled{2} Specification loss over long horizon.} Without explicit upfront elicitation, behavioral intent must be maintained implicitly across turns and is susceptible to drift as context grows~\cite{dongre2025drift, laban2025llms}. Compression mechanisms such as reasoning token stripping (i.e., removing the reasoning content between agent turns) and context summarization exacerbate this by selectively discarding tokens that encode the original specification~\cite{labate2025solving, ma2025quantifying}, so later-turn outputs increasingly reflect a degraded approximation of the target behavior.

\noindent\textbf{\circled{3} Error Propagation without anchoring.} Without an elicited specification as a stable reference, errors introduced early in the run accumulate and propagate across subsequent turns without correction. When the agent misinterprets the documentation or makes an incorrect behavioral assumption, it continues building on that assumption turn after turn---each dependent decision compounding the original error with no artifact to anchor against.

In traditional software engineering, these limitations are addressed by treating specification elicitation as a \emph{first-class phase} that precedes implementation~\cite{nuseibeh2000requirements, sommerville2016software}: requirements engineers exercise prototypes and construct scenarios to surface behavioral detail that written artifacts never capture, before a single line of code is written~\cite{wiegers2013software, zowghi2005survey}. We argue the same decomposition should apply to agentic program construction. Spec-driven frameworks such as OpenSpec~\cite{openspec2026} and spec-kit~\cite{speckit2026} adopt a similar philosophy, but rely solely on prompting human stakeholders for specification input, and do not support automated behavioral elicitation through binary probing.

We present \ourtool, a two-stage pipeline that explicitly separates behavioral specification elicitation from code synthesis. In the first stage, a dedicated \emph{spec agent} receives the documentation and execute-only binary and focuses exclusively on understanding the target program's behavior through systematic black-box probing, producing a structured behavioral specification artifact (i.e., \texttt{SPEC.md}). In the second stage, an existing \emph{code synthesis agent} receives the documentation, the binary, and \texttt{SPEC.md}, and focuses exclusively on implementation, now equipped with a behavioral model that goes beyond what the documentation alone provides. This decomposition directly addresses each limitation above: the spec agent explores freely without competing with coding, documentation ambiguities are resolved through probing before implementation begins, and the behavioral gap between documentation and binary is explicitly addressed rather than left to chance.

We evaluate \ourtool on the full 200-instance \textsc{ProgramBench} suite~\cite{yang2026programbench} across four models spanning two families and an order of magnitude of capability (Qwen3.5-397B-A17B, Qwen3.6-35B-A3B, GPT-5.5-high, and GPT-5.4-mini). Across all settings, \ourtool consistently outperforms the baseline (i.e., mini-SWE-agent),  with test pass rate improvement of 6.9\%--21.3\% and binary exploration coverage gains of 9.4\%--18.5\%, all statistically significant. Behavioral analysis further confirms that the decomposition restructures how the code synthesis agent allocates its budget: with a prior specification, the agent commits to writing code earlier and produces larger, more complete implementations.

This paper makes the following contributions:
\begin{enumerate}
  \item \textbf{\ourtool framework.}
    We propose the first agentic pipeline that explicitly decouples behavioral specification elicitation from code synthesis for from-scratch program construction, directly instantiating the requirements-engineering phase.

  \item \textbf{Empirical evaluation.}
    We evaluate on all 200 ProgramBench instances across four models, demonstrating consistent improvements in test pass rate (6.9\%--21.3\%) and binary exploration coverage (9.4\%--18.5\%), all statistically significant.

  \item \textbf{Behavioral analysis.}
   We show that having a prior specification changes how the code synthesis agent spends its turn budget: it starts writing code earlier and sustains implementation longer, rather than alternating between probing and coding throughout the run.
\end{enumerate}

The remainder of this paper is organized as follows.
Section~\ref{sec:motivation} presents the background and limitations of existing approaches. 
Section~\ref{sec:approach} introduces the details of \ourtool.
Section~\ref{sec:experimentalsetting} presents experimental settings. Section~\ref{sec:results} presents the evaluation and results. Section~\ref{sec:related} surveys the related work and discusses the differences between our work and previous studies. 
Section~\ref{sec:discussion} discusses failure cases, formats, costs, and threats to validity. 
Section~\ref{sec:conclusion} concludes our study.

\section{Background and Motivation}\label{sec:motivation}

\subsection{Problem Formulation}
\label{sec:approach:formulation}

Let $\mathcal{P}$ denote a target program that is a deterministic command-line program. The agent $\mathcal{L}$ observes $\mathcal{P}$ through two artifacts: (i) documentation $\mathcal{D}$ (e.g., a readme file) that describes $\mathcal{P}$ in natural language, and (ii) an execute-only binary $\mathcal{B}$ that instantiates $\mathcal{P}$'s observable behavior but whose source code and internal structure are inaccessible. The agent $\mathcal{L}$ task is to produce an implementation $\hat{\mathcal{I}}$ containing source code and necessary files to build an executable file, that reconstructs the behavior of $\mathcal{P}$. The implementation $\hat{\mathcal{I}}$ may be written in any programming language and architecture chosen by the agent. Next, it builds a candidate executable $\hat{\mathcal{B}}$ from $\hat{\mathcal{I}}$, and evaluates $\hat{\mathcal{B}}$ against a given hidden test suite $\mathcal{T} = \{t_i\}_{i=1}^{N}$, where each test $t_i = \bigl(\mathbf{a}_i,\, \mathbf{s}_i,\, \mathcal{B}(\mathbf{a}_i, \mathbf{s}_i)\bigr)$ consists of a command-line argument string $\mathbf{a}_i \in \Sigma^*$, a standard-input stream $\mathbf{s}_i \in \Sigma^*$, and the ground-truth output $\mathcal{B}(\mathbf{a}_i, \mathbf{s}_i)$ produced by $\mathcal{B}$ on that input. The objective is to produce $\hat{\mathcal{I}}$ that maximizes \emph{test pass rate}:
\begin{equation}
  \text{TestRate}(\hat{\mathcal{I}}, \mathcal{T})
  \;=\;
  \frac{1}{N}\sum_{i=1}^{N}
  \mathbf{1}\!\left[
    \hat{\mathcal{B}}(\mathbf{a}_i, \mathbf{s}_i)
    \;=\;
    \mathcal{B}(\mathbf{a}_i, \mathbf{s}_i)
  \right],
\end{equation}

which measures the fraction of test cases where $\hat{\mathcal{B}}$ reproduces the observable behavior of $\mathcal{P}$ exactly. A submission achieves $\text{PassRate} = 1.0$ if and only if $\hat{\mathcal{B}}$ is behaviorally equivalent to $\mathcal{B}$ on $\mathcal{T}$. Figure~\ref{fig:specAgent} (blue box) presents the general pipeline of existing code synthesis  agents~\cite{yang2024swe,wang2025openhands, yang2026programbench}. The agent operates in an iterative probe-then-build loop: it invokes $\mathcal{B}$ on selected inputs and reads $\mathcal{D}$ to build an understanding of the program's behavior, then progressively writes, tests, and refines source code until it judges the implementation sufficiently complete. The final output is a repository containing source code and a build file (e.g., \texttt{compile.sh}), which the evaluation harness compiles into a candidate executable $\hat{\mathcal{B}}$ and scores against hidden tests.

\subsection{Limitations of Existing Code Synthesis Agent}
\label{sec:limitation}

\begin{figure}[t]
\begin{lstlisting}[
  basicstyle=\scriptsize\ttfamily,
    frame=single,
    backgroundcolor=\color{gray!10},
    breaklines=true,
    breakatwhitespace=false,
    breakindent=0pt,
    columns=flexible,
    keepspaces=true,
    label={fig:readme},
    caption={\texttt{gomplate} README file
    ($\mathcal{D}$).}]
gomplate is a template renderer that supports multiple datasources (JSON, YAML, HTTP endpoints, environment variables, and more).

###Usage examples:

Basic usage with environment variables: echo 'Hello, {{ .Env.USER }}' | gomplate

Inline template evaluation: gomplate -i 'the answer is: {{ mul 6 7 }}'

Using a datasource file: gomplate -d config=./config.yaml -i 'the value we want is: {{ (ds "config").foo.bar.baz }}'

See the official documentation for full details and examples: https://docs.gomplate.ca

License: see the LICENSE file in this repository.


\end{lstlisting}
\end{figure}

In practice, $\mathcal{D}$ provides only a surface-level description of $\mathcal{P}$---primary features, typical usage, and representative examples---while remaining ambiguous or silent on edge cases, output formats, and error behavior~\cite{berry2004ambiguity, ferrari2014pragmatic}. Figure~\ref{fig:readme} shows the README for \texttt{gomplate}\footnote{\url{https://github.com/hairyhenderson/gomplate}}, a command-line template renderer in \textsc{ProgramBench}. The README describes the tool's purpose and lists common flags, but \texttt{gomplate} extends Go's \texttt{text/template} engine with a large library of built-in functions spanning namespaces such as \texttt{coll}, \texttt{math}, \texttt{net}, \texttt{crypto}, and \texttt{strings}. No README can enumerate the precise semantics, argument types, error conditions, and output formats of every function; faithful re-implementation therefore requires extensive behavioral exploration of the binary. We use this instance to illustrate the limitations of the existing single-loop framework.

\noindent\textbf{Ineffective exploratory probing.} Without a dedicated exploration phase, the baseline agent interleaves probing with coding, allocating only a fraction of its turns to behavioral discovery before committing to an implementation. For \texttt{gomplate}, this is particularly damaging: the agent must understand not only the top-level CLI interface but also the semantics of functions across ten namespaces, each with its own argument conventions and edge-case behavior. In practice, we observe that the general code synthesis agent (i.e., mini-SWE-agent using Qwen3.6-35B-A3B) confines its probing to the functions explicitly mentioned in the README---primarily the \texttt{env} and \texttt{data} namespaces---while leaving the behavior of \texttt{coll}, \texttt{math}, \texttt{net}, and \texttt{random} largely unverified.

\noindent\textbf{Specification loss over long horizon.}
Even when the collected specification is read correctly, its content can be lost across a long run if nothing anchors it. In \texttt{gomplate}, the general code synthesis agent (i.e., mini-SWE-agent using GPT-5.5-high) collected the full namespace index at turn 4, receiving function signatures for every namespace including \texttt{aws}, \texttt{gcp}, \texttt{sockaddr}, \texttt{semver}, \texttt{sprig}, and \texttt{cidr}. Yet across the remaining 25 turns, none of these six namespace names appears again in the agent's reasoning or tool calls. The information was observed once and never re-referenced: the final submission's function map registers 17 namespaces and silently omits all six, causing 126 test cases to fail before a single line of code is written for them.

\noindent\textbf{Error propagation without anchoring.}
Without a persistent specification, a single early misinterpretation can silently corrupt every subsequent implementation decision. In the \texttt{gomplate} instance under Qwen3.6-35B-A3B, the agent correctly reads the typed signature \texttt{Seq(start, end, step int64) []int64} from documentation at turns 4--5. Yet when \texttt{Seq} is first implemented at turn 57, it emits a degraded form---\texttt{func Seq(args ...interface{}) []interface{}}--- discarding both typed arguments and return value. Over the next 140 turns the math namespace is refactored five times, each time re-emitting the wrong signature without ever consulting the original documentation. The incorrect signature ships in the final submission, causing 25 of 43 \texttt{Seq}-related tests to fail with type errors.

These limitations motivate \ourtool's core design: a dedicated spec agent that systematically explores the specification before implementation begins, externalizing its findings into a persistent specification file (\texttt{Spec.md}) that eliminates the coverage gap from ineffective probing, prevents early misinterpretations from propagating uncorrected, and ensures that Behavioral knowledge discovered during exploration is never lost across turns.

\section{Approach of \ourtool}
\label{sec:approach}

\begin{figure*}
    \centering
    \includegraphics[width=1\linewidth]{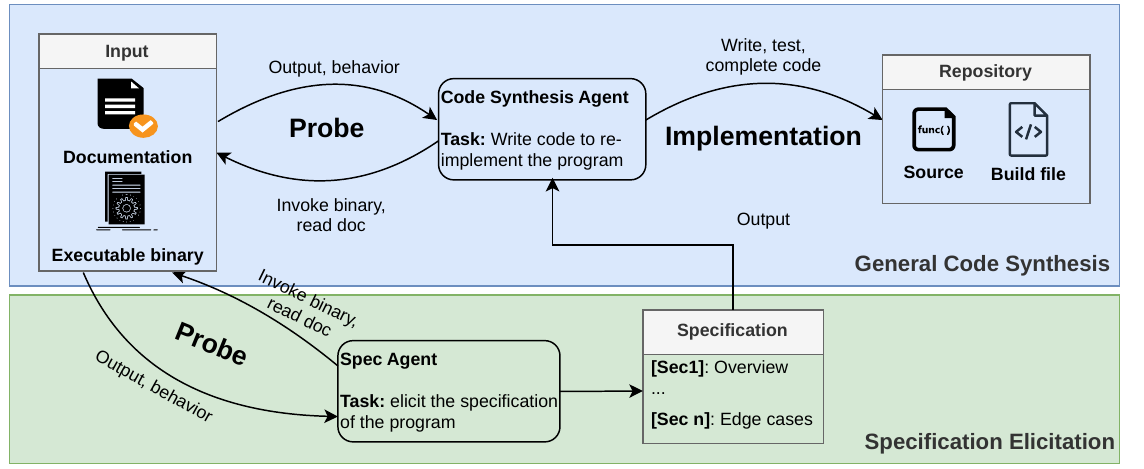}
    \caption{Framework of \ourtool. On top of the default code synthesis pipeline in the blue box, \ourtool introduces\textbf{spec agent} to elicit the specification before implementation in the green box.} 
    \label{fig:specAgent}
\end{figure*}

Our approach directly addresses all three limitations through a dedicated specification elicitation phase prior to implementation. We implement an autonomous \textbf{spec agent} that probes $\mathcal{B}$ systematically through black-box interaction, uncovering behaviors that documentation $\mathcal{D}$ leaves implicit or omits entirely --- directly countering \textbf{\circled{1}} by ensuring exploration is complete before coding begins. The elicited specification is persisted as an explicit artifact $\mathcal{S}$ (i.e., \texttt{SPEC.md} in our case), which serves as a stable behavioral contract that anchors all subsequent implementation decisions, resolving \textbf{\circled{2}} by providing an anchor that prevents early errors from compounding. Critically, because $\mathcal{S}$ is stored externally rather than maintained implicitly in the model's context, it remains intact regardless of context growth or compression, mitigating \textbf{\circled{3}}. The overall framework of \ourtool is presented in Figure~\ref{fig:specAgent}. The code synthesis agent can be any existing agent (e.g., OpenHands~\cite{wang2025openhands} or SWE-agent~\cite{yang2024swe}). If the agent encounters an ambiguity or gap in $\mathcal{S}$ during implementation, it may still probe binary $\mathcal{B}$ to resolve it. We elaborate on the design and operation of the spec agent in the following sections.

\subsection{Eliciting Behavior Specification through Probing of the Executable File}

At each turn, the spec agent selects a probe, i.e., an invocation of $\mathcal{B}$ with arguments, flags, or stdin chosen from what it has inferred so far, and observes the run time results from triple channels $(\texttt{stdout}, \texttt{stderr}, \texttt{exit})$. Free-form bash was chosen over a structured probe API because it mirrors how a human engineer would interact with an unfamiliar tool, and because it allows the agent to chain commands (e.g., constructing input files, running the binary, and inspecting its output in one step). For interactive or TUI programs, pre-installed \texttt{tmux}/\texttt{libtmux} provides a virtual terminal through which the agent can send keystrokes and capture screen state.

The spec agent conducts an iterative probing loop in which each observation of binary $\mathcal{B}$ informs the next probe. Starting from the behavioral skeleton provided by $\mathcal{D}$, the agent progressively deepens its exploration of $\mathcal{P}$ by targeting the classes of behavior that documentation characteristically under-specifies. Analyzing of the $\mathcal{D}$ reveals four recurring probing patterns that go beyond what $\mathcal{D}$ already provides:
\begin{itemize}
    \item \emph{\textbf{Boundary probing:}} testing inputs at the limits of accepted ranges (e.g., empty input, maximum-length strings, special characters) to determine exact boundary semantics that documentation leaves implicit. 
    \item \emph{\textbf{Error-path elicitation:}} systematically triggering error conditions (malformed input, missing required arguments, conflicting flags) to record the precise \texttt{stderr} message and exit code each condition produces. 
    \item \emph{\textbf{Combinatorial flag testing:}} exercising flags in combination to surface interaction effects absent from per-flag documentation. 
    \item \emph{\textbf{Output-format refinement:}} presenting adjacent inputs and comparing their outputs to resolve ambiguities in field ordering, delimiter choice, and whitespace handling that prose descriptions inevitably leave underspecified.
\end{itemize}

\subsection{Constraints and Shortcut Prevention}

We observe that LLM agents can shortcut the intended task in ways that inflate apparent performance without genuine behavioral understanding. Two shortcuts are particularly dangerous in this setting: (a) \emph{\textbf{source recovery}} locates the original source code via package registries, GitHub, or web search, which would trivialize re-implementation; (b) \emph{\textbf{binary introspection}} uses disassemblers, tracers, or decompilers to bypass black-box interaction entirely. Neither shortcut represents the specification-elicitation behavior we aim to study, and both would confound the results.

Therefore, to address this, when designing prompts, we explicitly prohibit LLM from doing both classes of actions.  All interactions with $\mathcal{B}$ must go through its normal user interface (CLI flags, stdin/stdout). An automated judge inspects the agent's command history for prohibited patterns (repository clones, registry installs, source tarball downloads, and disassembler calls); any detected violation disqualifies the run and records a score of zero.  Wrapping or shimming the provided binary is similarly prohibited and detected. This enforcement ensures that all behavioral knowledge encoded in the generated spec file derives from black-box observation, making the specification-elicitation loop the sole source of information beyond $\mathcal{D}$.

\subsection{Specification Deliverable and Format}\label{sec:format}
Even a thorough elicitation loop is only useful if its findings are communicated in a form the downstream exec agent can act on. The format of specification is a non-trivial design variable. A free-form transcript of raw observations is hard to act on, while an over-prescribed template may suppress behavioral detail that does not fit its categories. We adopt the \emph{sections} format: a light scaffold of six named headings: Overview; Flags; Input \& stdin; Output format; Error patterns; Edge cases. They are chosen as the minimal structure that covers the behavioral dimensions a re-implementer needs. The headings prescribe \emph{what} to document, not \emph{how} to discover or phrase it, ensuring the isolation between specification elicitation and code synthesis. More comparison of different formats (i.e., freeform and OpenSpec~\cite{openspec2026}) can be found in Section~\ref{sec:format}.

\subsection{Termination}
There is no semantic oracle for specification completeness: no external judge can assess whether \texttt{SPEC.md} captures all of $\mathcal{P}$'s observable behavior. An agent that terminates too early produces an incomplete spec; one that never terminates produces no spec at all. A run terminates under one of three conditions, applied in priority order.
\begin{itemize}
    \item \textbf{Self-declared completion (primary)}: if the spec agent judges the specification complete, it finalizes the process. Completeness is the agent's own judgment, which is itself a research variable we measure. 
    \item \textbf{Step limit}: if 1,000 agent--environment turns is reached, it terminated. The generous step limit (1,000 turns) is set to permit high-coverage exploration; in practice, most runs self-terminate well before this cap, confirming that the limit is a safety net rather than a binding constraint.
    \item \textbf{Wall-clock limit}: We use 6 hours as a final safeguard. The spec agent is terminated after running 6 hours.

\end{itemize}

Finally, a deliverable gate enforces artifact presence. If the specification file is absent at submission, the run is rejected and the agent is asked to write it (capped at 8 rejections, after which the run is recorded as a failure).

\subsection{Principles of Agent Design}

We introduce the principles of the agent design involved in \ourtool in this section.
\noindent\textbf{General agent architecture.} All our agents follow the ReAct agent design by following previous work~\cite{yang2025lingxi,yao2022react}. Except for that, we force all agents to emit reasoning output after each tool calling, including observation, considering alternative actions, and reasoning about next actions. This reasoning output forces explicit processing rather than blind tool call chaining and creates interpretable reasoning traces for debugging. \noindent\textbf{Context Management.} We compress information passing through the two stages (i.e., spec agent and code synthesis agent), to prevent information overload. More specifically, we condition the code synthesis agent with the structured behavioral specification artifact (i.e., SPEC.md), while discarding verbose reasoning and tool calling outputs from the spec agent. This preserves essential findings while removing redundant content that causes attention dilution when context windows become too large. We summarize the context information between stages rather than complete conversation histories~\cite{yang2024swe,yang2025lingxi}.

\section{Experimental Design}\label{sec:experimentalsetting}
In this section, we present our research questions (RQs),
studied dataset, retrievers, used prompt templates, implementation details, and our approach to RQs.

\subsection{Research Question}
We aim to answer the following research question:
\begin{itemize}
    \item \textbf{RQ1: What is the effectiveness of \ourtool?}
     \item \textbf{RQ2: How effective is \ourtool across task difficulty levels?}
     \item \textbf{RQ3: Does the Spec Agent improve behavioral exploration coverage?}
    \item \textbf{RQ4: How does \ourtool affect the behavior of the code synthesis agent?}
\end{itemize}

In RQ1, we evaluate the effectiveness of \ourtool in synthesizing programs from scratch and compare it with selected baseline. In RQ2, we evaluate the effectiveness of \ourtool across instances with different difficulty levels. In RQ3, we evaluate the specification exploration ability of \ourtool and compare it with the baseline. In RQ4, we investigate how the spec agent affects the behavior of the code synthesis agent by comparing pipeline configurations with and without spec elicitation.

\subsection{Benchmark}
\label{sec:eval:benchmark}

We evaluate on the full \textsc{ProgramBench} suite~\cite{yang2026programbench}, which comprises 200 command-line program instances drawn from real-world open-source tools, covering widely used software and complex applications, such as FFmpeg, SQLite, and the PHP interpreter. Those 200 task instances in ProgramBench are built from repositories written primarily in compiled languages: Rust (107), Go (46 repositories), C/C++ (45 repositories), 1 (Java), and 1 (Haskell). Each instance provides a natural-language documentation (a README and man page) describing the program, and an execute-only binary that serves as the behavioral oracle during spec elicitation and implementation. Each instance is labelled by difficulty---easy (28), medium (143), and hard (29)---and annotated with the implementation language of the reference binary. To evaluate the correctness and coverage of the implemented program, the benchmark contains 248,853 test functions across its 200 tasks, averaging a median of 770 tests per task. These suites achieve an average of 79.7\% line coverage, and are hidden from the code synthesis agent.

\subsection{Evaluation Metric}
\noindent\emph{\textbf{Average test pass rate}} as the primary metric. For each instance $p$, let $k_p$ denote the number of hidden test cases passed by the candidate executable and $n_p$ the total number of test cases for that instance. The instance-level pass rate is $r_p = k_p / n_p$, and the benchmark-level
score is the mean over all instances:
\begin{equation}
  \bar{r}
  \;=\;
  \frac{1}{|\mathcal{P}|}\sum_{p \in \mathcal{P}} \frac{k_p}{n_p}.
\end{equation}
This metric awards partial credit for implementations that reproduce a subset of the target's behavior, making it sensitive to incremental improvements that a binary resolved/unresolved rate would mask. Note that even the strongest baseline, GPT-5.5-high without spec elicitation, fully resolves only 1 of 200 instances (resolved rate 0.5\%)~\cite{yang2026programbench}, yet achieves a non-trivial $\bar{r}$ reflecting partial behavioral coverage across many instances. Average test pass rate therefore provides a meaningful signal for comparing pipeline configurations in a regime where full resolution is rare.

\noindent\textbf{Probing Coverage.}
To quantitatively evaluate the specification exploration ability of our approach against the baseline, we introduce \textit{probing coverage} as the primary metric. Given a target binary executable $\mathcal{B}$, let $\mathcal{C} = \{c_1, c_2, \ldots, c_N\}$ denote the set of all $N$ executable lines of code in $\mathcal{B}$. During the probing phase, the agent interacts with $\mathcal{B}$ by issuing a sequence of probes; for each probe, we instrument $\mathcal{B}$ to record the set of lines executed. We use a language-specific instrumentation tool for each language. For instance, we use \texttt{go build -cover} for Go binaries, \texttt{gcc --coverage} for C/C++ binaries, and \texttt{cargo llvm-cov} for Rust binaries to capture line-level execution traces for every probe issued during the session. Upon completion of the probing phase, we aggregate the unique recorded lines across all probes to obtain the covered set $\mathcal{C}_{\text{probed}} \subseteq \mathcal{C}$. Probing coverage is then defined as:

\begin{equation}
    \text{Coverage} = \frac{|\mathcal{C}_{\text{probed}}|}{|\mathcal{C}|} \times 100\%
    \label{eq:probing_coverage}
\end{equation}

\noindent A higher probing coverage indicates that the agent has exercised a larger portion of the program's behavior during specification, reflecting a more thorough and complete understanding of the binary's functionality.

\subsection{Models}
\label{sec:eval:models}

We evaluate four models spanning two families and an order of magnitude of capability, selected to disentangle the effect of model scale from the effect of model family. Table~\ref{tab:models} summarizes their key properties. All four models are evaluated with reasoning enabled, reflecting the multi-step inference the task demands. The Qwen models expose no effort-level control and are evaluated with thinking on by default. We do not tune any other decoding parameters and use provider defaults throughout, to avoid conflating model capability with hyper-parameter optimization.

\begin{table}[t]
  \centering
  \caption{Models used in the evaluation.}
  \label{tab:models}
  \resizebox{1\columnwidth}{!}{%
  \begin{tabular}{l|c|c|c}
    \hline
    \textbf{Model} & \textbf{Family} & \textbf{Type} & \textbf{Reasoning} \\
    \hline
    \textbf{Qwen3.5-397B-A17B}~\cite{yang2025qwen3}   & Open-weights & MoE   & On (default)  \\
    \textbf{Qwen3.6-35B-A3B}~\cite{yang2025qwen3}     & Open-weights & MoE   & On (default)  \\
    \textbf{GPT-5.5-high}~\cite{openai2026gpt55}      & Proprietary  & Unknown & High effort   \\
    \textbf{GPT-5.4-mini}~\cite{openai2026gpt54}      & Proprietary  & Unknown & Medium effort \\

    \hline
  \end{tabular}%
  }
\end{table}

\subsection{Baselines}

\textbf{\baselineDirect}:
The baseline is the original \textsc{ProgramBench} pipeline~\cite{yang2026programbench}, in which the code synthesis agent receives $(\mathcal{D}, \mathcal{B})$ and proceeds directly to implementation \textit{without} any preceding specification elicitation phase. This baseline establishes the performance ceiling of the standard approach and measures the net gain attributable to the entire specification elicitation.

Note that we use the same scaffold (i.e., official mini-swe-agent) as the code synthesis agent for both \baselineDirect and \ourtool. In other words, the only difference between \baselineDirect and \ourtool is that we introduce spec agent in \ourtool, and other settings are kept the same.

\subsection{Implementation details}

Each run executes in an isolated, no-network Docker container from the task image by ProgramBench.  The same model is used for both specification elicitation and code synthesis phases. Our local modifications to each are released as patches against those exact commits. All models are accessed through a single OpenAI-API-compatible routing proxy LiteLLM~\cite{litellm2023}, with a fresh context per pipeline phase to prevent information leakage between the spec and exec runtimes. All experiments were conducted using model APIs accessed via a stable API provider.

\section{Results}\label{sec:results}

\subsection{RQ1 - Effectiveness of \ourtool}\label{sec:rq1}

\subsubsection{Approach}
We compare \ourtool against \baselineDirect across all four models, with all other experimental conditions held constant. For each model, we run both configurations on the full 200-instance suite and compute the per-instance pass rate $r_p$ for each instance. To assess whether the improvement in average pass rate is statistically significant, we apply a paired two-sided Wilcoxon signed-rank test ($\alpha = 0.05$).  We report win/loss/tie counts to characterise the direction and prevalence of the effect across instances. To further understand whether \ourtool improves not just the average performance, but the proportion of high-quality implementations, we report the fraction of instances achieving a pass rate above a threshold $t$ (i.e., 50\% - 95\%) for both \ourtool and \baselineDirect. 

\begin{table}[t]
  \centering
  \caption{Comparison of \ourtool and \baselineDirect on average test pass rate
           (\%).            W/L/T counts the number of instances where \ourtool outperforms, underperforms, and ties \baselineDirect, respectively. We present the improvement of \ourtool over \baselineDirect in ().}
  \label{tab:rq1}
  \resizebox{1\columnwidth}{!}{%
  \begin{tabular}{l|c|c|c}
    \hline
    \textbf{Model} & \textbf{\baselineDirect} & \textbf{\ourtool} & \textbf{W/L/T} \\
    \hline
    \textbf{Qwen3.5-397B}  & 33.66\% & 40.84\% \textbf{(+21.3\%)} & 130/58/12 \\
    \textbf{Qwen3.6-35B}    & 27.51\% & 31.40\% \textbf{(+14.1\%)} & 121/67/12 \\
    \textbf{GPT-5.5-high}       & 59.02\% & 65.14\% \textbf{(+10.4\%)} & 150/38/12 \\
    \textbf{GPT-5.4-mini}       & 39.09\% & 41.78\% \textbf{(+6.9\%)}  & 119/69/12 \\
    \hline
  \end{tabular}
  }
\end{table}

\subsubsection{Results}
\textbf{\ourtool consistently improves the average test
pass rate across all four models, with relative gains ranging from 6.9\% to 21.3\%, up to 65.14\% with GPT-5.5-high.} 
Table~\ref{tab:rq1} shows that \ourtool consistently improves the average test pass rate across all four models, with relative gains ranging from 6.9\% (GPT-5.4-mini) to 21.3\% (Qwen3.5-397B-A17B). All improvements are statistically significant ($p < 0.01$), confirming that the gains are not attributable to random variation. The win/loss/tie counts reinforce this conclusion: across all models, \ourtool outperforms the baseline at least on more than 59\% of instances, with the strongest model (GPT-5.5-high) winning on 75\% (150/200) instances. Taken together, these results confirm that explicitly eliciting a behavioral specification before implementation is a robust and model-agnostic intervention that materially improves program re-implementation fidelity.

\textbf{\ourtool does not merely lift average pass rates; it significantly expands the upper tail of the score distribution by driving a higher proportion of programs to near-perfection (i.e., the programs that pass almost all tests).} 
Table~\ref{tab:hit_rate_threshold} presents the proportion of programs, where the test pass rate is larger than a threshold  (50\% to 95\%). For instance, at the most stringent thresholds, \ourtool triples the rate of near-perfect solutions from 5.5\% to 16.5\% (for $t \ge 90\%$) and more than quadruples it from 1.5\% to 6.5\% (for $t \ge 95\%$). We observe a consistently similar trend across the other evaluated models.

\begin{table}[h]
\centering
\caption{Proportion of programs that have a test pass rate larger than a threshold $t$ using GPT-5.5-high, with the 95\% confidence interval in \textbf{[]}.}
\label{tab:hit_rate_threshold}
\begin{tabular}{l|c|c}
\hline
\textbf{$\ge t (\%)$} & \textbf{Baseline} & \textbf{\ourtool} \\ \hline
$\ge 50$          & 69.0\% [62.3\%, 75.0\%]    & 74.0\% [67.5\%, 79.6\%]  \\
$\ge 70$          & 42.5\% [35.9\%, 49.4\%]    & 55.5\% [48.6\%, 62.2\%]  \\
$\ge 90$          & 5.5\% [3.1\%, 9.6\%]       & 16.5\% [12.0\%, 22.3\%]  \\
$\ge 95$          & 1.5\% [0.5\%, 4.3\%]       & 6.5\% [3.8\%, 10.8\%]    \\ \hline
\end{tabular}
\vspace{-0.1in}
\end{table}

\begin{figure}[t]
\begin{lstlisting}[
    basicstyle=\scriptsize\ttfamily,
    frame=single,
    backgroundcolor=\color{gray!10},
    breaklines=true,
    breakindent=0pt,
    columns=flexible,
    keepspaces=true,
    label={fig:spec_excerpt},
    caption={Selected entries from \texttt{SPEC.md} elicited by
    \textsc{SpecAgent} for the \texttt{gomplate} instance.}]

## Function Set
The executable provides gomplate namespaces and aliases in addition to Go template built-ins. Important observed functions and aliases include:

- Data: datasource/ds, include, datasourceExists, datasourceReachable...
- Collections: coll.Dict/dict, coll.Slice, coll.GoSlice...

## Flags
$ ./executable --help
Process text files with Go templates
Usage:
  gomplate [flags]
Flags:
  -d, --datasource datasource     datasource in alias=URL form.   Specify multiple times to add multiple sources.
  ...

## Edge Cases and Compatibility Notes
  --in plus --file or --input-dir: invalid (exclusive-input conflict).
  --file plus --input-dir: likewise invalid. 
  ...

\end{lstlisting}
\end{figure}

Listing~\ref{fig:spec_excerpt} shows an example specification elicited for the \texttt{gomplate} instance.  For example, \ourtool records a thorough \texttt{SPEC.md} containing all function sets and edge cases. An exec agent that reads it before writing code can implement each of these behaviors correctly on the first attempt rather than guessing, which directly reduces the behavioral residual between $\hat{\mathcal{B}}$ and $\mathcal{B}$ and explains the consistent pass rate lift across all four models.

\subsection{RQ2 - Effectiveness across different difficulty levels}\label{sec:rq2}

\subsubsection{Approach}
To assess the robustness of \ourtool, we analyze its performance across different task-difficulty tiers: Easy ($n=28$), Medium ($n=143$), and Hard ($n=29$), with the difficulty tags provided in the benchmark.

\begin{table*}[ht]
\centering
\caption{Comparison of the average test pass rates between Direct-Synthesis and \ourtool across task difficulty levels. Cells show the pass rate change as Direct-Synthesis $\rightarrow$ \ourtool (relative improvement).}
\label{tab:baseline_specagent_difficulty}
\begin{tabular}{l|c|c|c}
\hline
\textbf{Model} & \textbf{Easy ($n=28$)} & \textbf{Medium ($n=143$)} & \textbf{Hard ($n=29$)} \\
\hline
\textbf{Qwen3.5-397B} & 42.9\% $\rightarrow$ 52.9\% (+23.3\%) & 34.6\% $\rightarrow$ 42.1\% (+21.7\%) & 20.0\% $\rightarrow$ 23.0\% (+15.0\%) \\
\textbf{Qwen3.6-35B} & 37.8\% $\rightarrow$ 42.2\% (+11.6\%) & 28.1\% $\rightarrow$ 32.1\% (+14.2\%) & 14.5\% $\rightarrow$ 17.6\% (+21.4\%) \\ 
\textbf{GPT-5.5-high} & 73.7\% $\rightarrow$ 78.9\% (+7.1\%) & 61.9\% $\rightarrow$ 67.5\% (+9.0\%) & 30.8\% $\rightarrow$ 40.0\% (+29.9\%) \\
\textbf{GPT-5.4-mini} & 49.7\% $\rightarrow$ 57.4\% (+15.5\%) & 40.7\% $\rightarrow$ 42.8\% (+5.2\%) & 21.0\% $\rightarrow$ 21.9\% (+4.3\%) \\
\hline
\end{tabular}
\end{table*}

\subsubsection{Result}
\textbf{\ourtool consistently outperforms Direct-Synthesis across all difficulty levels and all evaluated models. For instance, GPT-5.5-high experiences its most pronounced surge, improving the average test pass rate from 30.8\% to 40.0\%, which represents a remarkable 29.9\%.} 
Rather than being restricted to a specific tier of tasks or a particular model architecture, the \ourtool shows a universal improvement.  More importantly, the performance gains are not confined to trivial or moderately difficult tasks. \ourtool delivers substantial improvements on the hardest problems. Crucially, the largest absolute improvements often manifest where the baseline struggles the most. For instance, on Hard instances, the frontier model GPT-5.5-high experiences its most pronounced surge, improving the average test pass rate from 30.8\% to 40.0\%, which represents a remarkable 29.9\% improvement. This pattern indicates that a rigorous, decoupled specification phase serves as a vital cognitive scaffold, allowing models to systematically decompose intricate logic that would otherwise cause a direct code-generation attempt to fail.

\subsection{RQ3 - Effectiveness of Spec Agent}

\subsubsection{Approach} The spec agent is dedicated to probing the binary and eliciting a behavioral specification before implementation begins. To assess whether it actually improves exploratory coverage, we measure probing coverage of \ourtool and compare it against \baselineDirect{}, which performs no dedicated
elicitation phase.

\subsubsection{Results}

\begin{table}[ht]
\centering
\caption{Probing coverage of \ourtool versus \baselineDirect. For \ourtool, coverage is reported separately for the spec agent, the code synthesis agent, and their union. Win denotes the proportion of instances where \ourtool's union coverage exceeds \baselineDirect.}
\label{tab:two_steps_separately}
\begin{tabular}{l|c|c|c}
\hline
\textbf{Model} & \textbf{\ourtool} & \textbf{\baselineDirect} & \textbf{Win} \\ 
 & Spec / Synthesis / Union & & \\ \hline
Qwen3.5-397B & 58.0\%/51.0\%/\textbf{59.8\%} & 51.9\% & 77.2\%\\
Qwen3.6-35B   & 54.9\%/51.3\%/\textbf{58.3\%} & 49.2\% & 71.2\%\\
GPT-5.5-high      & 58.3\%/31.2\%/\textbf{60.3\%} & 55.1\% & 66.2\% \\
GPT-5.4-mini      & 56.5\%/41.7\%/\textbf{58.6\%} & 52.1\% & 63.5\%\\ \hline
\end{tabular}
\end{table}

\textbf{\ourtool consistently elicits broader behavioral coverage than \baselineDirect significantly, achieving 9.4\%--18.5\% improvement. The gain originates from the spec agent rather than from the code synthesis agent.} 
Table~\ref{tab:two_steps_separately} reports probing coverage across all four models. \ourtool achieves 58.3\%--60.3\% total coverage, an improvement of 9.4\%--18.5\% over \baselineDirect on every model. All improvements are statistically significant ($p < 0.05$). Its spec agent covers more behaviors than the baseline in 63.5\%--77.2\% of instances.

Decomposing \ourtool by component reveals that the spec agent alone (54.9\%--58.3\%) consistently exceeds the code synthesis agent (31.2\%--51.3\%) on all models, and the union adds only marginally over the spec agent. This confirms that the coverage gain is driven by the dedicated elicitation phase, not by incidental exploration during code synthesis.

\subsection{RQ4 - Impact of Spec Agent on Code Synthesis Agent's behavior}\label{sec:rq4}

\subsubsection{Approach} 
To understand how the elicited specification shapes the code synthesis agent's behavior during the run, we record the codebase size (lines of code) at every agent--environment turn for both \ourtool and \baselineDirect, and normalise each turn index to $[0, 1]$ by dividing by the run's total number of turns. 

\begin{figure*}
    \centering
    \includegraphics[width=1\linewidth]{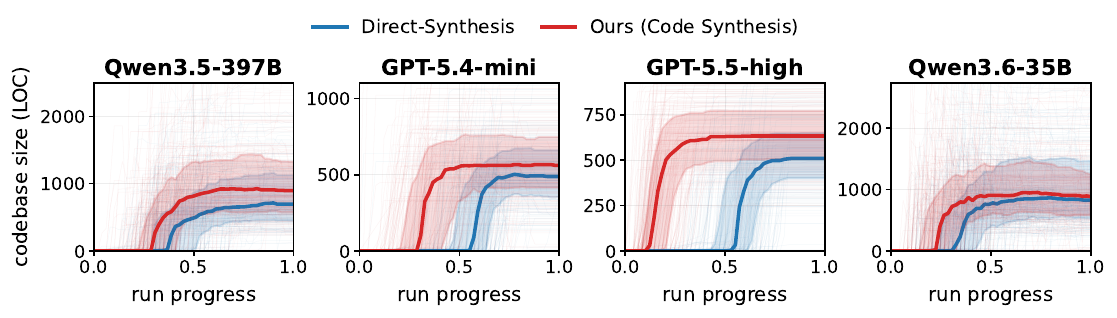}
    \caption{Median reconstructed codebase size (lines of code) as a function of normalised run progress (0 = start, 1 = termination), with shaded interquartile range (IQR) across instances. The $y$-axis is capped at the 95th percentile of final codebase sizes to suppress outliers. A curve that rises earlier indicates the agent commits to writing code sooner in its budget; a higher terminal value indicates a larger final codebase. } 
    \label{fig:behavior}
\end{figure*}

\subsubsection{Results}
\begin{figure}
    \centering
    \includegraphics[width=1\linewidth]{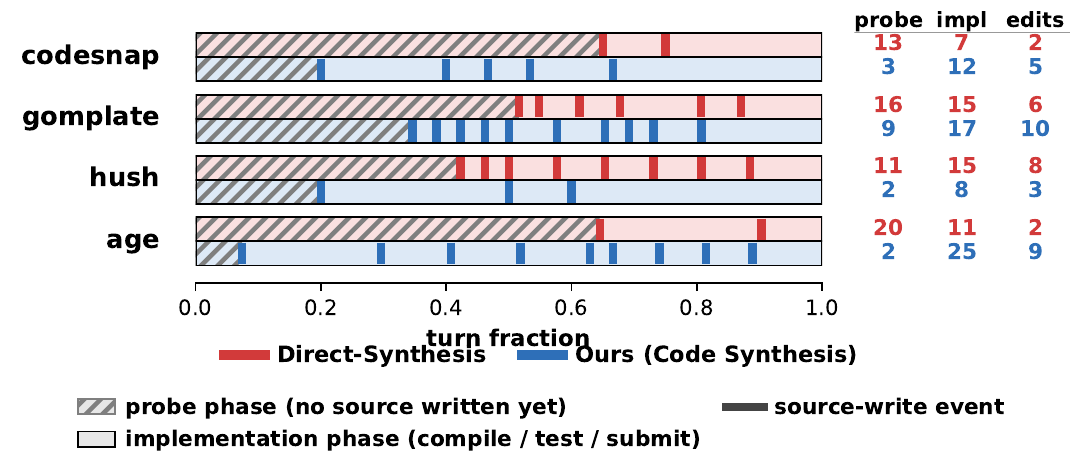}
    \caption{Comparison of the code synthesis phase between \ourtool and baseline from four GPT-5.5 cases. Hatched areas represent the exploration calls, and solid areas denote the implementation calls. Vertical ticks indicate source-write turns.}
    \label{fig:exampleBehavior}
\end{figure}

\textbf{Providing a persistent behavioral specification before implementation shifts the code synthesis agent from a probe-then-build pattern to an earlier, more sustained building phase.} 
Figure~\ref{fig:behavior} tracks the growth of the reconstructed codebase as the code synthesis agent progresses through its turns. \ourtool starts implementation earlier than baseline. In the baseline, the early portion of the run is dominated by exploratory probing. For instance, Figure~\ref{fig:exampleBehavior} compares the probing phase and implementation phase of \ourtool and the baseline on four instances. On \texttt{age}, for example, the baseline spends most turns inspecting behavior before attempting to dump the entire 841-line implementation in a single action, inevitably failing due to an exhausted debugging budget. The baseline typically exhausts 11 to 20 turns probing the binary's behavior. In contrast, \ourtool directly resolves this bottleneck by externalizing exploration into a persistent \texttt{SPEC.md} during the dedicated spec phase. Instead of wasting implementation turns on trial-and-error probing, the code synthesis agent spends just 2 to 9 turns establishing context before immediately shifting to implementation.

\textbf{\baselineDirect agents stop early, not because they run out of budget, but because they run out of behavioral understanding.} 
One possible attribute of the weaker performance of \baselineDirect to insufficient turn budget: if probing consumes too many turns, the agent has little budget left for implementation. To rule out this explanation, we examine the number of turns consumed by \baselineDirect. The results are striking: zero instances reach the 1,000-turn step limit, and all instances terminate early by agent choice, consuming a median of only 22--177 turns (2--18\% of the available budget). The bottleneck is therefore not turn budget but the quality of behavioral understanding the agent constructs before it stops: agents terminate voluntarily because they believe they have understood enough, not because they exhausted their budget. In other words, simply allocating more compute would not help. \ourtool addresses this by equipping the code synthesis agent with a persistent specification established through the dedicated spec agent prior to implementation, enabling higher coverage of the specification, as demonstrated in RQ3.

\textbf{Introducing the spec agent leads to larger codebases and higher pass rates across all models.}
The final codebase produced under \ourtool is 7--29\% larger across models (shaded IQR), consistent with agents writing more complete implementations rather than terminating early under behavioral uncertainty. This aligns with RQ3, where we demonstrate that a higher probing coverage from the spec agent yields a more complete specification, which in turn drives broader implementation coverage.

Together, these results show that specification elicitation does not merely provide additional context---it fundamentally restructures how the exec agent allocates its effort, shifting the run profile from a probe-then-build pattern toward an earlier, more sustained building phase.

\section{Related Work}\label{sec:related}

\subsection{LLM-based Code Generation and Program Synthesis}
\label{sec:related:codegen}

The application of large language models to code synthesis has progressed from template-filling approaches~\cite{gulwani2017programsynthesis} through neural program induction~\cite{devlin2017robustfill} to contemporary autoregressive models trained on hundreds of billions of tokens of source code~\cite{chen2021codex,nijkamp2022codegen,roziere2023codellama, guo2024deepseek,li2023starcoder}. These models achieve impressive performance on function-level benchmarks such as HumanEval~\cite{chen2021codex} and MBPP~\cite{austin2021mbpp}, where a docstring and a handful of unit-test examples provide a highly constrained, unambiguous specification. Scaling the same paradigm to repository-level or whole-program tasks reveals a pronounced degradation. SWE-bench~\cite{yang2024swe} demonstrates that resolving real GitHub issues, which requires understanding repository context far harder than synthesizing isolated functions. Agentic frameworks~\cite{wang2024openhands,yang2024swe,gao2025trae,yang2025lingxi} couple LLMs with tool-use loops (shell, file editors, web search) to extend their reach,  yet they still rely on a task description written by a human that is treated as a fixed, complete specification. \textsc{ProgramBench}~\cite{yang2026programbench} introduces a harder variant: the reference behavior is embodied in a black-box binary, not a text description, forcing the agent to \emph{discover} the specification rather than consume it. Our work is the first to explicitly design the discovery step as a first-class, reusable pipeline component when building programs from scratch. 

\subsection{Requirements Elicitation in Software Engineering}
\label{sec:related:re}

Requirements engineering (RE) is among the most studied disciplines in SE~\cite{sommerville2016software,wiegers2013software,nuseibeh2000requirements}, with a recurring finding that written artifacts systematically omit behavioral detail, contain ambiguities, and conflict with one another~\cite{berry2004ambiguity,ferrari2014pragmatic}, making elicitation through interaction the standard remedy~\cite{canedo2022creativity,zowghi2005survey}. NLP and LLM techniques have been applied to automate parts of this process, including ambiguity detection~\cite{ferrari2014pragmatic}, traceability recovery~\cite{hayes2006advancing}, and requirements inconsistency detection~\cite{bashir2025requirements}, but these approaches treat requirements as textual artifacts to be \emph{processed} rather than behaviors to be \emph{discovered} through interaction. \textsc{SpecAgent} takes the complementary view, instantiating an LLM as an active elicitor that probes a running system rather than analyzing static documents---closest in spirit to behavior-driven development~\cite{smart2023bdd,karhu2009exploratory}, but producing a requirements artifact rather than a test suite.

\subsection{Black-box Program Analysis and Specification Extraction}
The closest prior work to \ourtool's elicitation loop is the line of research on protocol reverse engineering, which probes network binaries to extract message formats and state machines~\cite{caballero2007polyglot,comparetti2009prospex}. These approaches share our black-box observation model but produce machine-readable protocol grammars for security analysis rather than natural-language behavioral specifications for code generation. Specification mining~\cite{ammons2002mining} and dynamic invariant detection~\cite{ernst2007daikon} infer formal properties from execution traces, but require either source instrumentation or passive observation of existing traces rather than active, agent-directed probing. LLM-based decompilation~\cite{tan2024llm4decompile} recovers source from binaries at the instruction level and requires disassembly access that our setting explicitly prohibits. Program translation  approaches~\cite{roziere2020transcoder} operate source-to-source and assume a complete, readable specification is already available. \ourtool occupies a distinct position: it actively probes a black-box binary under maximal access restrictions and produces a natural-language behavioral specification consumable directly by a downstream LLM code synthesis agent.

\section{Discussion}\label{sec:discussion}

\subsection{Failure case analysis}\label{sec:rq4}

To understand the failure case and identify their root causes, we randomly sample 50 test failure cases across four models and manually label each by examining three artifacts jointly: (i) the failing-test message (what behavior the test exercises and what concrete output, exit code, or error string it expects), (ii) the Spec.md (whether and how precisely the behavior is described), and (iii) the relevant source snippet (whether the implementation contains the corresponding code path and whether it matches Spec.md). We observed the following five types of failures:

\noindent\textbf{F1: Spec Omission (10\%).}
  Spec.md never documents the features that the test exercises (e.g., the CLI flag, subcommand, error path, or output channel). It means that the spec agent failed to discover and collect the behavior from the binary.

\noindent\textbf{F2: Wrong Spec Description (4\%).}
  Spec.md describes the behavior incorrectly (e.g., attributing a message to \texttt{stderr} when the test asserts \texttt{stdout}). The implementation was faithful to Spec.md, but Spec.md itself contradicted the binary's actual behavior.

\noindent\textbf{F3: Inaccurate Spec Description (26\%).}
    Spec.md documents the feature under test, but the description is insufficiently precise, forcing the implementation agent to guess. For example, Spec.md may document that invalid input produces an error without specifying the exit code or output channel; the implementation chooses \texttt{rc=1} with a \texttt{stderr} message, while the test expects \texttt{rc=2} with a \texttt{stdout} message.
    
\noindent\textbf{F4: Execution Fault (52\%).}
  Spec.md describes the behavior correctly and completely, yet the implementation diverges. Common shapes include a subcommand recognized in help text but rejected by the dispatcher, a flag was parsed but its effect never applied, or an error format that contradicts Spec.md's documented convention. This is the dominant failure mode, indicating that even a correct Spec.md does not guarantee a correct implementation for complex
  behavioral constraints.

\noindent\textbf{F5: Environment-dependent Error (8\%).}
  Failures are environment-dependent (e.g., NS server strings embedded in error messages, missing test-wrapper paths, or subprocess timeouts from absent toolchains).  These reflect benchmark infrastructure noise rather than agent deficiencies.

\noindent\textbf{Implications.}
Although \ourtool prove the use of spec agent is a promising direction for building a program from scratch. Our failure analysis suggests the following research direction. The dominant failure mode (F4, 52\%) indicates that closing the remaining gap requires stronger execution-stage reasoning (e.g., better instruction following, self-consistency checking against Spec.md, and test-driven repair loops), rather than improvements to spec elicitation alone.  F1, F2, and F3 (40\% combined) suggest that elicitation coverage and behavioral fidelity remains an open problem; future work could further explore targeted probing strategies that systematically exercise under-documented surfaces such as error paths, flag interactions, and output routing.

\subsection{Does the specification's format matter?}\label{sec:format}

To understand the impact of specification format, we compared the following formats, differing only in the deliverable-instruction paragraph of the task prompt; all other prompt text, probing rules, and limits are identical across conditions. Note that all the results reported in RQs are based on Sections.

\noindent\textbf{Sections.}
    The format we used is introduced in Section~\ref{sec:format}, where the agent is asked to write \texttt{SPEC.md} as structured Markdown under named headings.

\noindent\textbf{Freeform.}
    The agent chooses its own structure and coverage criteria. This condition isolates whether imposed structure helps or hurts downstream exec quality.

\noindent\textbf{OpenSpec.}
    The agent writes \texttt{SPEC.md} in the OpenSpec format~\cite{openspec2026}, structuring the document as a \texttt{\#\# Purpose} section followed by \texttt{\#\#\# Requirement:} blocks, each annotated with RFC~2119 keywords (\textsc{must}/\textsc{shall}, \textsc{should}, \textsc{may}), and concrete \texttt{\#\#\#\# Scenario:} examples in GIVEN/WHEN/THEN form. This condition tests whether a more formal, requirements-engineering vocabulary improves the precision of the elicited specification.

We compare the average test pass rate of \ourtool using the above different format with GPT-5.4-mini on randomly sampled 50 instances\footnote{We choose to use GPT-5.4-mini and a subset due to budget limit.}. Table~\ref{tab:spec_format_comparison} compares the effectiveness of different formats. As shown, every specification format improves on the baseline, while Sections improves the most and outperforms other formats. 

\begin{table}[htbp]
\centering
\caption{Specification Format Comparison}
\label{tab:spec_format_comparison}
\begin{tabular}{lc}
\hline
\textbf{Format} & \textbf{Pass Rate (\%)} \\ \hline
\textbf{Baseline} (No Spec) & 55.9\% \\
\textbf{Freeform} & 60.7\% \\
\textbf{OpenSpec} & 61.7\% \\
\textbf{Sections} & \textbf{62.6\%} \\ \hline
\end{tabular}
\vspace{-0.1in}
\end{table}

\subsection{Cost of \ourtool}

\begin{table}[ht]
\centering
\caption{Mean per-instance cost (USD) for \baselineDirect{} and \ourtool{}. For \ourtool{}, cost is broken down into the spec agent, code synthesis agent, and their total; percentages indicate overhead relative to \baselineDirect{}.}
\label{tab:mean_cost}
\resizebox{\columnwidth}{!}{%
\begin{tabular}{l|c|c}
\hline
\textbf{Model} & \textbf{\baselineDirect} & \textbf{\ourtool~(Spec/Synthesis/Total)} \\ \hline
Qwen3.5-397B & \$2.56 & \$0.95 / \$2.84 (+11\%) / \$3.79 (+48\%) \\
Qwen3.6-35B   & \$2.03 & \$0.50 / \$2.68 (+32\%) / \$3.17 (+56\%) \\ 
GPT-5.4-mini      & \$0.35 & \$0.25 / \$0.29 (-17\%) / \$0.53 (+51\%) \\
GPT-5.5-high      & \$2.54 & \$3.16 / \$2.69 (+6\%) / \$5.85 (+130\%) \\
\hline
\end{tabular}%
}
\end{table}

Table~\ref{tab:mean_cost} reports the mean per-instance cost for \baselineDirect{} and \ourtool{}. The total cost of \ourtool{} is higher by 48\%--130\% across models, which is expected: the spec agent constitutes an additional inference phase (\$0.25--\$3.16 per instance), and the more complete specification produced leads the code synthesis agent to generate larger, more feature-complete implementations, modestly increasing its own cost as well (+6\%--+32\% for most models). The exception is GPT-5.4-mini, where synthesis cost decreases by 17\%, suggesting that a clearer upfront specification reduces exploratory overhead during coding. The largest overhead occurs with GPT-5.5-high (+130\%), driven primarily by the cost of the spec agent itself rather than by synthesis. Overall, the cost increase is a direct reflection of \ourtool doing more useful work---broader behavioral exploration and more complete implementation---rather than redundant computation.

\subsection{Threats to Validity}

\noindent\textbf{Internal Validity } The \ourtool uses two agents instead of one, meaning it spends more budget than the baseline (without spec agent). The risk is that the improvement is not actually from the value of having a specification, but simply because the model got a larger compute budget. To alleviate this threat, as shown in RQ4, 100\% of the runs ended early by choice. Therefore, giving the baseline models more compute wouldn't have naturally increased their test pass rate.

\noindent\textbf{External Validity}
Our evaluation covers deterministic command-line programs in \textsc{ProgramBench}; results may not generalize to programs with non-deterministic behavior, graphical interfaces, or complex inter-process communication. We evaluate four models on a single scaffold (SWE-agent); while consistent gains across two model families suggest robustness, results may differ for other models and scaffolds. Although \ourtool requires an executable binary for spec elicitation, the binary serves as a proxy for any \emph{executable oracle} that exposes a probe-and-observe interface, such as a running REST API, a containerized service, a compiled SDK, or a remote CLI are all equivalent from the spec agent's perspective. The key prerequisite is not a binary specifically but \emph{any queryable reference or human}, a condition met by most software maintenance, migration, and integration scenarios.

\noindent\textbf{Construct Validity} The study's main evaluation metric (a 0–100\% scale average test pass rate) combines two different things: whether the agent can actually build a working program, and whether that program behaves correctly. There was a risk that failures such as compilation failures (like a missing compile.sh) would unfairly tank the average scores. However, these total technical failures were extremely rare (at most 3 out of 200 runs). This means when an agent scored a 0\%, it was not because it could not build a working program; it was because the program's actual behavior was completely wrong.

\section{Conclusion}\label{sec:conclusion}
We presented \ourtool, a two-stage agentic framework that explicitly decouples behavioral specification elicitation from code synthesis for from-scratch program re-implementation. By dedicating a separate spec agent to systematically probe the binary and produce a structured specification prior to implementation, \ourtool addresses three fundamental limitations of single-loop baselines: insufficient behavioral exploration, specification loss over multi-turns, and uncorrected error propagation. Experiments on all 200 \textsc{ProgramBench} instances across four frontier models demonstrate consistent improvements in test pass rates (6.9\%--21.3\%) and binary exploration coverage (9.4\%--18.5\%). Our results suggest that treating requirements elicitation as a first-class phase is a promising paradigm for advancing LLM agents on complex, from-scratch software construction tasks.

\newpage

\bibliographystyle{IEEEtran}
\balance

\bibliography{references}
\end{document}